\documentstyle[12pt]{article} 
\topmargin
-1.5cm 
\begin{document} 
\scrollmode

\vglue 1.7cm

\noindent {\large 
{\bf On a Generalized Oscillator System: Interbasis Expansions}}

\vspace{2.5cm}
\noindent {\bf MAURICE KIBLER}

\noindent {Institut de Physique Nucl\'eaire de Lyon, IN2P3-CNRS et Universit\'e 
Claude Bernard,}\\
{43 Bd du 11 Novembre 1918, F-69622 Villeurbanne Cedex, France}

\vspace{0.8cm}
\noindent {\bf LEVON G.~MARDOYAN}

\noindent {Laboratory of Nuclear Problems,}\\
{Joint Institute for Nuclear Research, 141980 Dubna, Moscow region, Russia}

\vspace{0.8cm}
\noindent {\bf GEORGE S.~POGOSYAN}

\noindent {Bogoliubov Laboratory of Theoretical Physics,}\\
{Joint Institute for Nuclear Research, 141980 Dubna, Moscow region, Russia}

   \setcounter{page}{0}

\vspace{2cm}
\noindent {\bf ABSTRACT}

\noindent 
This article deals with a nonrelativistic quantum mechanical
study of a dynamical system which generalizes the isotropic
harmonic oscillator system in three dimensions.
The Schr\"odinger equation for this generalized
oscillator system is separable in spherical, cylindrical,
and spheroidal (prolate and oblate) coordinates. The
quantum mechanical spectrum
of this system is worked out in some details. The problem of interbasis
expansions of the wavefunctions is completely solved. The coefficients
for the expansion of the cylindrical basis in terms of the spherical
basis, and vice-versa, are found to be analytic continuations
(to real values of their
arguments) of Clebsch-Gordan coefficients
for the group SU(2).
The interbasis expansion coefficients for the prolate and
oblate spheroidal bases in terms of the spherical or the cylindrical
bases are shown to satisfy three-term recursion relations.
Finally, a connection between the generalized oscillator system
(projected on the $z$-line) and the Morse system (in one dimension) 
is discussed. 

\vspace{2cm}


   \newpage


\vspace{0.3cm} 
{\bf Introduction}
\vspace{0.1cm}

The purpose of this paper is to study the quantum mechanical motion of 
a particle in the three-dimensional axially symmetric potential 
\begin{equation}
V= \frac{\Omega^2}{2} (x^2+y^2+z^2) 
 + \frac{P}       {2} \frac{1}{z^2} 
 + \frac{Q}       {2} \frac{1}{x^2+y^2},
\label{eq:K1}
\end{equation}
where $\Omega$, $P$, and $Q$ are constants with
      $\Omega > 0$, $P > - \frac{1}{4}$, and $Q \geq 0$. In the last decade, 
this potential (including the case $P=0$) has
been the object of numerous studies [1-12].
The Schr\"odinger and Hamilton-Jacobi equations 
for this generalized
oscillator potential are separable in spherical, cylindrical, 
and spheroidal (prolate and oblate) coordinates. In the case when 
$P=0$ we get the well-known ring-shape
oscillator potential which was investigated in many articles 
\cite{Q1,LMPS4,KW9,GGZ35,KLW5} 
in recent years as a companion of the Hartmann potential 
               [4, 6, 8, 13-17]. 
If $P=Q=0$ we have the ordinary 
isotropic harmonic oscillator in three dimensions.

The plan of this article is as follows. We 
solve the Schr\"odinger equation 
\begin{equation}
H \Psi = E \Psi,
\label{eq:K2}
\end{equation}
with 
\begin{equation}
H = - \frac{1}{2} 
(\partial_{xx}  + 
 \partial_{yy}  + 
 \partial_{zz}) + V,
\nonumber 
\end{equation}
in spherical   coordinates (in the second section) and 
in cylindrical coordinates (in the third  section). 
(The constant $\hbar$ and the reduced mass 
 are taken to be equal to $1$. 
In the whole paper, we use $\Psi$ to denote
the total wavefunction whatsoever the coordinate system is; 
the wavefunctions $\Psi$ in spherical, cylindrical, and 
spheroidal coordinates are then distinguished by the
corresponding quantum numbers. 
Note also that we use $s$ to denote the 
fraction $\frac{1}{2}$ in the following.) 
In the fourth section, we determine the interbasis expansion 
coefficients between the cylindrical and the spherical bases. 
The fifth and sixth sections deal 
with the generalized oscillator system 
in spheroidal coordinates.
In particular, the prolate and oblate
spheroidal bases are expanded in terms of both the spherical basis
and the cylindrical basis. Two appendices close this article.
The first  one      is devoted to the bi-orthogonality  of the radial 
wavefunctions (in spherical coordinates) for the generalized 
oscillator system. 
The second appendix concerns a connection between the
Smorodinsky-Winternitz system (that is a basic component for
the generalized oscillator system) and the Morse system. 

The generalized oscillator system 
constitutes a pending part to the generalized Kepler-Coulomb 
system studied in Ref.~\cite{Z7,GPS8,KMP27}. The latter two 
nonrelativistic systems 
generalize two important paradigms in
quantum mechanics, namely, the oscillator system and 
the Kepler-Coulomb system. The main results of this work and the
one in Ref.~\cite{KMP27} concern the separability in spheroidal coodinates 
as well as the SU(2) approach and the three-term recursion relations for the 
interbasis expansion coefficients. 

The authors are very pleased to contribute to this memorial volume in honour of
Jean-Louis Calais. Professor Jean-Louis Calais achieved, among other 
important works, an original job \cite{JLC39} 
on the derivation of the SU(2) Clebsch-Gordan
coefficients by the (L\"owdin) projection operator method. We are glad to
present here a work where an analytic continuation of SU(2) Clebsch-Gordan
coefficients plays an important r\^ole in the analysis of interbasis
expansions. 

The use of spheroidal coordinates is now well established in 
quantum chemistry \cite{PC40}. There exist now powerful techniques 
\cite{ABRAMOV} for evaluating (angular and radial) prolate spheroidal
wavefunctions from differential equations. It is hoped that this paper 
will shed some new light on expansions of spheroidal wavefunctions. 

\vspace{0.3cm} 
{\bf Spherical Basis}
\vspace{0.1cm}

The Schr\"odinger equation (\ref{eq:K2}) in spherical coordinates 
$(r, \theta, \varphi)$ 
              for the potential (\ref{eq:K1}), i.e., 
\begin{eqnarray} 
V= \frac{ \Omega^2 }{2} r^2 
 + \frac{P}         {2} \frac{1}{r^2\cos^2\theta} 
 + \frac{Q}         {2} \frac{1}{r^2\sin^2\theta},
\nonumber 
\end{eqnarray}
may be solved by seeking a 
wavefunction $\Psi$ of the form
\begin{equation}
\Psi(r,\theta,\varphi) =
R(r)\Theta(\theta)\frac{{\rm e}^{i m\varphi}}{\sqrt{2\pi}},
\label{eq:K4}
\end{equation}
with $m \in {\bf Z}$. This amounts to find the eigenfunctions of the set 
$\{H,L_z,M\}$ of commuting 
operators, where the constant of motion 
$M$ reads 
\begin{eqnarray}
M = L^2 + \frac{P}{\cos^2\theta} 
        + \frac{Q}{\sin^2\theta} 
\label{eq:K34}
\end{eqnarray}
($L^2$ is the square of the angular 
momentum and ${L_z}$ its $z$-component). We are thus left with 
the system of coupled differential equations:
\begin{eqnarray}
(M - A) \Theta &=& 0, 
\label{eq:K5}
\\
\left[ \frac{1}{r^2} d_r ( r^2 d_r ) 
+ 2E - \Omega^2 r^2 - \frac{A}{r^2} \right] R &=& 0,
\label{eq:K6}
\end{eqnarray}
where $A$ is a (spherical) separation constant. 

Let us consider the angular equation (\ref{eq:K5}). By putting 
$\Theta(\theta) = f(\theta) / \sqrt{\sin\theta}$, 
we can rewrite Eq.~(\ref{eq:K5}) in the P\"oschl-Teller form:
\begin{eqnarray}
\left( d_{\theta\theta} 
+  A 
+ \frac{1}{4} 
-\frac{b^2 - \frac{1}{4}}{\cos^2\theta}
-\frac{c^2 - \frac{1}{4}}{\sin^2\theta} \right) f = 0,
\nonumber
\\
\label{eq:K8}
\\     
      b = \sqrt{P + \frac{1}{4}}, 
\quad c = \sqrt{Q + m^2}.
\nonumber 
\end{eqnarray}
In the case where $b > s$, the angular 
potential is repulsive for $\theta = \frac{\pi}{2}$. 
In this case,  the $\theta$ domain is separated in two regions 
($\theta \in \ ]0, \frac{\pi}{2}  [$ and 
 $\theta \in \ ]\frac{\pi}{2}, \pi[$) 
and the ``motion'' takes place in one or another region. 
Furthermore, in  this case Eq.~(\ref{eq:K8}) corresponds to a genuine 
P\"oschl-Teller potential. In the case where $0 < b < s$, 
we can call the angular potential an attractive P\"oschl-Teller 
potential. When $b = s$, i.e., $P=0$, 
we get the well-known ring-shape oscillator 
potential \cite{Q1,LMPS4,KW9,GGZ35,KLW5}. 
The solution $\Theta(\theta) \equiv \Theta_{q}(\theta; c,\pm b)$ 
of Eq.~(\ref{eq:K5}) (for both $0 < b < s$ and 
                                   $b > s$), 
with the conditions $\Theta(0) = \Theta(\frac{\pi}{2}) = 0$, 
is easily found to be (cf., \cite{E123,FL24})
\begin{equation}
\Theta(\theta) = 
N_{q}(c, \pm b)(\sin\theta)^{c}(\cos\theta)^{s\pm b}
P_q^{(c,\pm b)}(\cos2\theta), 
\label{eq:K11}
\end{equation}
with $q \in {\bf N}$, where $P_{n}^{(\alpha,\beta)}$ 
denotes a Jacobi polynomial. Then, the constant $A$ is quantized as
\begin{equation}
A_q(c, \pm b) = (2q + c \pm b +  s )
                (2q + c \pm b + 3s ).
\label{eq:K12}
\end{equation}
The normalization constant $N_{q} (c, \pm b)$ in (\ref{eq:K11}) is
given (up to a phase factor) by
\begin{eqnarray}
\int_{0}^{\frac{\pi}{2}}
\Theta_{q'} 
\Theta_{q } 
\sin\theta d{\theta}=
\frac{1}{2}\delta_{q'q}. 
\label{eq:K13}
\end{eqnarray}
This leads to 
\begin{eqnarray}
N_{q}(c, \pm b)
= \sqrt\frac{(2q+c \pm b+1)q!\Gamma(q+ c\pm b+1)}
{\Gamma(q+c
+1)\Gamma(q \pm b+1)}.
\label{eq:K14}
\end{eqnarray}
Note that only the positive sign in front of 
$b$ has to be taken when $b > s$
while both the positive and negative signs have to be considered for 
$0 < b < s$. 

Let us go to the radial equation (\ref{eq:K6}). The introduction
                              of (\ref{eq:K12}) into 
                                 (\ref{eq:K6}) yields an equation 
that is very reminiscent of the radial equation for the three-dimensional
isotropic oscillator except that the orbital quantum number $l$ is 
replaced by $2q + c \pm b + s$. The solution 
$R(r) \equiv  R_{n_r q}(r; c, \pm b)$ of the obtained 
equation, in terms of Laguerre polynomials $L_n^{\alpha}$, is
\begin{equation}
R(r) = 
N_{n_r q}(c, \pm b)({\sqrt\Omega}r)^{ 2q + c \pm b + s }
{\rm e}^{-s\Omega{r^2}}L_{n_r}^{2q+ c \pm b+1}(\Omega{r^2}),
\label{eq:K16}
\end{equation}
with $n_r \in {\bf N}$. In Eq.~(\ref{eq:K16}), 
the radial wavefunctions $R_{n_r q}$ satisfy the 
orthogonality relation
\begin{equation}
\int_{0}^{\infty}
R_{n_r^{'} q}
R_{n_r     q}
r^2dr = \delta_{n_r^{'} n_r}
\label{eq:K17}
\end{equation}
[cf., Eq.~(\ref{eq:K135})] so
that the normalization factor $N_{n_r q}(c, \pm b)$ is 
\begin{equation}
N_{n_r q}(c, \pm b) = \sqrt{\frac{2\Omega^{3s} n_r! }
{\Gamma(n_r + 2q+ c \pm b +2)}}.
\label{eq:K18}
\end{equation}

The normalized total wavefunction 
$\Psi(r,\theta,\varphi) \equiv 
 \Psi_{n_r q m}(r,\theta,\varphi; c, \pm b)$ is then given by
          Eqs.~(\ref{eq:K4}),  
               (\ref{eq:K11}), 
               (\ref{eq:K14}),
               (\ref{eq:K16}), 
           and (\ref{eq:K18}). 
The energies $E$ corresponding to $n_r + q$ fixed are 
\begin{eqnarray}
E_n (c, \pm b) = \Omega \, (2n + c \pm b + 2), 
\label{eq:K19}
\end{eqnarray}
with $n = n_r + q$. 
Equation (\ref{eq:K19}) shows that, 
for each quantum number $n$, we have two levels (for $+ b$ and $- b$) 
 in the $0 < b < s$ region and one level (for $+ b$) in the 
            $b > s$ region. 
Note that this spectrum was 
obtained through a path integral approach in 
\cite{CB2,C3} for the $ b > s $ case and in 
\cite{GPS8} for the             general case (see also Refs.~\cite{KC6,Z7}). 

In the $0 < b < s$ region, for the limiting situation where 
$b=s^{-}$, i.e., $P = 0^{-}$, 
we have for the separation constant $A$: 
\begin{eqnarray}
A_{q}(c, + s ) = (2q+c+1)(2q+c+2), 
\nonumber
\\
\label{eq:K20}
\\ 
A_{q}(c, - s ) = (2q+c  )(2q+c+1).
\nonumber 
\end{eqnarray}
Then, by using the connecting formulas \cite{BE13}
\begin{eqnarray}
C_{2n+1}^{\lambda}(x) = \frac{(\lambda)_{n+1}}{(s)_{n+1}}
\, x \, P_n^{(\lambda-s,+s)}(2x^2-1), 
\nonumber
\\
C_{2n  }^{\lambda}(x) = \frac{(\lambda)_{n  }}{(s)_{n  }}
     \, P_n^{(\lambda-s,-s)}(2x^2-1),
\nonumber 
\end{eqnarray}
between the Jacobi polynomial $P_n^{(\alpha,\beta)}$ and the
Gegenbauer polynomial $C_n^{\lambda}$, we have 
the following odd and even 
angular solutions (with respect to $\cos\theta \mapsto 
                                   -\cos\theta$)
\begin{eqnarray}
\Theta_{q} (\theta; c,+s) = 
\sqrt{\frac{(4q + 2c + 3)(2q+1)!}{{2\pi}\Gamma(2q+2c+2)}}
\nonumber
\\
\times 2^{c}\Gamma (c+s) (\sin\theta)^{c}C_{2q+1}^{c+s}(\cos\theta),
\label{eq:K24}
\\
\Theta_{q} (\theta; c,-s) = 
\sqrt{\frac{(4q + 2c + 1)(2q)!}{2\pi\Gamma(2q+2c+1)}}
\nonumber
\\
\times 2^{c}\Gamma (c+s) (\sin\theta)^{c}C_{2q}^{c+s}(\cos\theta).
\label{eq:K25}
\end{eqnarray}
Let us introduce (a new orbital quantum number) $l$ 
and (a new principial quantum number) $N$ through 
   \begin{eqnarray}
   l - |m| = \left \{ \matrix {
           2q + 1  & {\rm for \ the \ + \ sign} \cr
     \hfill    2q  & {\rm for \ the \ - \ sign} \cr
          } \right \}, 
\nonumber 
\\
\label{eq:K26} 
\\ 
   N - |m| = \left \{ \matrix {
           2n + 1  & {\rm for \ the \ + \ sign} \cr
     \hfill    2n  & {\rm for \ the \ - \ sign} \cr
          } \right \}. 
\nonumber   
   \end{eqnarray}
Note that $N = 2n_r + l$ both for the $+$ and $-$ signs. 
Then, the separation constant [Eq.~(\ref{eq:K20})] and the 
                       energy [Eq.~(\ref{eq:K19})] can be
expressed as 
\begin{eqnarray}
A_{q} (c, \pm s) \equiv 
A_{l}(\delta) = (l+\delta)(l+\delta+1), 
\nonumber
\\
\label{eq:K28}
\\
E_{n} (c, \pm s) \equiv E_{N}(\delta)
= \Omega \, (N + \delta + 3 s),
\nonumber 
\end{eqnarray}
respectively, where 
\begin{eqnarray}
\delta = \sqrt{Q + m^2} - |m|.
\nonumber 
\end{eqnarray}
Thus, the two parts of the energy spectrum 
for the signs $\pm$ 
correspond now to odd (for $+$) 
             and even (for $-$)
values of $N-|m|$. In terms of $N$, $l$, and $\delta$, the functions
$R_{n_r q} (r; c, \pm s) \equiv R_{N l}(r; \delta)$ 
[cf., Eq.~(\ref{eq:K16})] and  
$\Theta_{q} (\theta; c, \pm s) \equiv \Theta_{lm}(\theta; \delta)$ 
[cf., Eqs.~(\ref{eq:K24}) and (\ref{eq:K25})] can
be rewritten as 
\begin{eqnarray}
R_{N l}(r; \delta) = \sqrt{\frac{2\Omega^{3s}(\frac{N - l}{2})!}
{\Gamma(\frac{N+l}{2} +\delta+3s)}}
\nonumber
\\
\times ({\sqrt\Omega}r)^{l+\delta}
{\rm e}^{-s\Omega{r^2}}
L_{\frac{N-l}{2}}^{l + \delta + s}(\Omega{r^2}),
\label{eq:K30bis}
\end{eqnarray}
\begin{eqnarray}
\Theta_{lm}(\theta; \delta) =
2^{|m|+\delta} \Gamma (|m|+\delta+s)
\nonumber
\\
\times \sqrt \frac{(2l+2\delta+1)(l-|m|)!}
{2\pi\Gamma(l+|m|+2\delta+1)}
\nonumber 
\\
\times (\sin\theta)^{|m|+\delta}C_{l-|m|}^{|m| + \delta + s}(\cos\theta). 
\label{eq:K30}
\end{eqnarray}
Equations (\ref{eq:K30bis}) and 
          (\ref{eq:K30} compare with the corresponding results for the 
ring-shape oscillator in \cite{Q1,LMPS4}. 
Note that (\ref{eq:K30}) was given in terms of Legendre functions 
in Refs.~\cite{Q1} and \cite{LMPS4} and was studied 
in details in Ref.~\cite{MST14}.

In the $b > s$ region, for the limiting situation where
$b = s^+$, i.e., $P = 0^+$, we have only odd solutions. 
In other words when $P \to 0^+$, the eigenvalues 
and eigenfunctions of the 
generalized oscillator do not restrict to the
eigenvalues and eigenfunctions, respectively, of the ring-shape 
oscillator. This fact may be explained in the following manner. 
To make $P=0$ in the wavefunction 
$\Psi_{n_r q m}(r,\theta,\varphi; c, + b)$ 
amounts to changing the Hamiltonian into a 
Hamiltonian corresponding to $P = 0$ and to 
introducing an unpenetrable barrier. 
(Another way to describe this phenomenon is to 
say that for very small $P$, the potential 
$V$ is infinite in the $\theta = \frac{\pi}{2}$ plan 
and equal to the ring-shape potential only for $P=0$.) 
This phenomenon is known as the Klauder phenomenon \cite{KL25}.

A further limit can be obtained in the case when $\delta = 0$, 
i.e., $Q=0$. It is enough to use the connecting formula
\cite{BE13}
\begin{eqnarray}
P_l^{|m|}(x)=\frac{(-2)^{|m|}}{\sqrt\pi}
\Gamma (|m|+s) (1-x^2)^{s|m|} C_{l-|m|}^{ |m| + s }(x)
\nonumber 
\end{eqnarray}
between the 
Gegenbauer polynomial $C_n^{\lambda}$ and the Legendre polynomial 
$P_l^{|m|}$. In fact for $Q=0$, Eq.~(\ref{eq:K30}) can be reduced to
\begin{eqnarray}
\Theta_{lm}(\theta; 0) = 
(-1)^{|m|}\sqrt{ \frac{2l+1}{2} \frac{(l-|m|)!}{(l+|m|)!} }
P_l^{|m|}(\cos\theta),
\nonumber
\end{eqnarray}
so that $\Theta_{lm}(\theta; 0){\rm e}^{{\rm i}m\varphi}/\sqrt{2\pi}$
coincides with the usual spherical harmonic $Y_{lm}(\theta,\varphi)$ 
(up to a phase factor, e.g., see \cite{VMK15}). The wavefunctions 
$\Theta_{q}(\theta; c,\pm b) { {\rm e}^{i m\varphi} } / {\sqrt{2\pi}}$ 
may thus be considered as a generalisation of the spherical harmonics. 

\vspace{0.3cm} 
{\bf Cylindrical Basis}
\vspace{0.1cm}

In the cylindrical coordinates
($\rho, \varphi, z$), the potential $V$ reads
\begin{eqnarray}
V = \frac{ {\Omega}^2 }{2} ({\rho}^2+z^2) + 
    \frac{P}{2}             \frac{1}{z^2} +
    \frac{Q}{2}        \frac{1}{{\rho}^2}.
\nonumber
\end{eqnarray}
Equation (\ref{eq:K2}), with this potential, 
admits a solution $\Psi$ of the form
\begin{equation}
\Psi(\rho,\varphi,z)=R(\rho) Z(z) \frac{{\rm e}^{im\varphi}}{\sqrt{2\pi}},
\label{eq:K37}
\end{equation}
where $m \in {\bf Z}$. 
In other words, we look for the eigenfunctions of the set 
$\{H,L_z,N\}$ of commuting 
operators, where the constant of motion $N$ is 
\begin{equation}
N = D_{zz} + \frac{P}{z^2}, 
\label{eq:K52}
\end{equation}
$D_{zz}$ being the $zz$ component of 
\begin{eqnarray}
D_{z_i z_j} = - \partial_{z_i z_j} + \Omega^2 \, z_i z_j,
\nonumber
\end{eqnarray}
the so-called Demkov tensor \cite{D19} 
for the isotropic harmonic oscillator in ${\bf R}^3$. 
It is sufficient to solve the two coupled equations
\begin{eqnarray}
( N - 2E_z ) Z &=& 0,
\label{eq:K38bis}
\\ 
\left[ \frac{1}{\rho} d_{\rho} ( \rho d_{ \rho} ) + 
2E_{\rho}-{\Omega}^2{\rho}^2-\frac{Q + m^2}{\rho^2} \right] R &=& 0, 
\label{eq:K38}
\end{eqnarray}
where the two cylindrical separation constants $E_{\rho}$ and 
$E_z$ obey $E_{\rho}+E_z=E$. 
The solutions 
$\Psi(\rho,\varphi,z) \equiv
 \Psi_{n_{\rho} p m}(\rho,\varphi,z; c, \pm b)$ 
of (\ref{eq:K38bis}-\ref{eq:K38}) 
lead to the normalized wavefunction 
\begin{equation}
\Psi(\rho,\varphi,z) =
R_{n_{\rho}}(\rho; c) Z_p(z; \pm b) \frac{{\rm e}^{im\varphi}}
{\sqrt{2\pi}},
\label{eq:K40}
\end{equation}
where
\begin{eqnarray}
R_{n_{\rho}}(\rho; c) = \sqrt{\frac{2\Omega n_{\rho}! }
{\Gamma(n_{\rho}+ c+1)}}{\rm e}^{ -s \Omega {\rho}^2 }
(\sqrt{\Omega}\rho)^{c}L_{n_{\rho}}^{c}(\Omega{\rho}^2) 
\nonumber
\end{eqnarray}
and 
\begin{eqnarray}
{Z}_p(z; \pm b) = (-1)^{p}\sqrt{\frac{ {\Omega}^{s} p!}
{\Gamma(p\pm b+1)}}
{\rm e}^{ -s \Omega z^2 }(\sqrt{\Omega}z)^{s\pm b}
L_p^{\pm b}(\Omega z^2), 
\label{eq:K42}
\end{eqnarray}
with $n_{\rho} \in {\bf N}$ and $p \in {\bf N}$. 
The normalization of the wavefunction (\ref{eq:K40}) is ensured by
\begin{eqnarray}
\int_{0}^{\infty}
R_{n'_{\rho}}
R_{n _{\rho}}
\rho d{\rho}={\delta}_{n'_{\rho}n_{\rho}}, 
\nonumber 
\\
\int_{0}^{\infty}
{Z}_{p'}
{Z}_{p }
dz = \frac{1}{2}{\delta}_{p'p}.
\nonumber
\end{eqnarray}
Furthermore, the constants $E_{\rho}$ and $E_z$ 
in (\ref{eq:K38bis}-\ref{eq:K38}) become
\begin{eqnarray}
E_{\rho}(n_\rho, c) = \Omega \, (2n_{\rho}+ c+1), 
\nonumber
\\
\label{eq:K45}
\\
E_z (p, \pm b) = \Omega \, (2p \pm b+1).
\nonumber
\end{eqnarray}
Therefore, the quantized values of the energy $E$ are given by 
(\ref{eq:K19}) where now the quantum number $n$ is $n=n_{\rho}+p$. 
As in the second section, the sign in front of $b$ in 
Eqs.~(\ref{eq:K40})-(\ref{eq:K45}) may be only positive when 
$b > s$. When $0 < b <s$, both the signs
$+$ and $-$ are admissible. 

In the $0 < b < s$  region, in the limiting 
case where $b=s^{-}$, due to the connecting 
formulas \cite{BE13}
\begin{eqnarray}
{\cal H}_{2n+1}(x) = (-1)^n 2^{2n+1} n! x L_n^{+s}(x^2),
\nonumber
\\ 
{\cal H}_{2n  }(x) = (-1)^n 2^{2n  } n!   L_n^{-s}(x^2),
\nonumber
\end{eqnarray}
between the odd  ${\cal H}_{2n + 1} $
        and even ${\cal H}_{2n}     $ Hermite polynomials and the
Laguerre polynomials $L_n^{\pm s}$, we immediately have
\begin{eqnarray}
Z_{p} (z; +s) &=&
\left(\frac{\Omega}{\pi}\right)^{\frac{1}{4}}
\frac{{\rm e}^{ -s {\Omega z^2} }}
{\sqrt{2^{2p+1}(2p+1)!}}
{\cal H}_{2p+1}(\sqrt{\Omega}z), 
\nonumber 
\\
Z_{p} (z; -s) &=&
\left(\frac{\Omega}{\pi}\right)^{\frac{1}{4}}
\frac{{\rm e}^{ -s {\Omega z^2} }}
{\sqrt{2^{2p}(2p)!}}
{\cal H}_{2p}(\sqrt{\Omega}z). 
\nonumber 
\end{eqnarray}
Introducing (a new quantum number) $n_3$ such that 
$n_3 = 2p+1$ for the $+$ sign and 
$n_3 = 2p  $ for the $-$ sign, 
we obtain
\begin{eqnarray}
Z_{p} (z; \pm s)
= 
\left(\frac{\Omega}{\pi}\right)^{\frac{1}{4}}
\frac{{\rm e}^{ -s {\Omega z^2} }}
{\sqrt{2^{n_3} n_3! }}
{\cal H}_{n_3}(\sqrt{\Omega}z).
\nonumber
\end{eqnarray}
The energy is then given by (\ref{eq:K28}) where $N = 2n_\rho + n_3 + |m|$. 
Note that the spectrum in the case $b = s^{-}$, which 
corresponds to the ring-shape oscillator system, was obtained in
  Refs.~[1-3].

In the $b >s$ region, in the limiting situation where
$b = s^{+}$, we get only the odd solution of 
the ring-shape oscillator system. 

\vspace{0.3cm} 
{\bf Connecting the Cylindrical and Spherical Bases}
\vspace{0.1cm}

According to first principles, any cylindrical 
wavefunction (\ref{eq:K37}) corresponding to a given value 
of $E$ can be developed in terms of the spherical wavefunctions 
             (\ref{eq:K4}) associated to the eigenvalue $E$ 
(see also Ref.~\cite{Z7}). Thus, we have 
\begin{equation}
  {\Psi}_{n_{\rho} p m}                                =
\sum_{q=0}^n W_{np}^q(c, \pm b)
  {\Psi}_{n_r      q m}                                ,
\label{eq:K54}
\end{equation}
where $n_\rho + p = n_r + q = n$. In Eq.~(\ref{eq:K54}), it is 
understood that the wavefunctions in the left- 
and right-hand sides are written in spherical
coordinates $(r,\theta,\varphi)$ owing to 
$\rho                       = r \sin \theta$ 
and 
$z                          = r \cos \theta$.
The dependence on ${\rm e}^{{\rm i} m \varphi}$ can be
eliminated in both sides of Eq.~(\ref{eq:K54}). Furthermore, by 
using the formula 
$L_n^{\alpha}(x) \sim (-1)^n {x^n} / {n!}$, 
valid for $x$ arbitrarily large, (\ref{eq:K54}) yields an equation
that depends only on the variable $\theta$. 
Thus, by using the orthonormality relation (\ref{eq:K13}), for the
quantum numbers $q$, we can derive the following expression 
for the interbasis expansion coefficients
  \begin{eqnarray}
                 W_{np}^q (c, \pm b)
 = (-1)^{q-p} \, B_{np}^q (c, \pm b) \, 
                 E_{np}^q (c, \pm b),
  \label{eq:K57}
  \end{eqnarray}
where
  \begin{eqnarray}
  B_{np}^q (c, \pm b) =
  \sqrt{\frac{(2q+ c \pm b + 1)
  (n-q)!q!\Gamma(q+ c \pm b+1)
  \Gamma(n + q + c \pm b + 2)}
  {(n-p)!p!\Gamma(q+ c+1)\Gamma(q\pm b+1)
  \Gamma(n-p + c + 1)\Gamma(p \pm b+1)}}
  \nonumber
  \\ 
  \label{eq:K59}
  \\
  E_{n p}^{q}(c, \pm b) =
  2 \,\int_{0}^{\frac{\pi}{2}}(\sin\theta)^{2n - 2p + 2c}
  (\cos\theta)^{2p+1\pm 2b}\, P_{q}^{(c, \pm b)}
  (\cos 2\theta) \sin\theta d \theta. 
  \nonumber 
\end{eqnarray}
By making the change of variable $x = \cos 2 \theta$ and 
by using the Rodrigues formula for the Jacobi polynomial 
\cite{BE13} 
\begin{eqnarray}
P_n^{(\alpha,\beta)}(x)=\frac{(-1)^n}{2^nn!}
(1-x)^{-\alpha}
(1+x)^{-\beta }\frac{d^n}{dx^n}[(1-x)^{\alpha + n}
                                (1+x)^{\beta  + n}],
\nonumber 
\end{eqnarray}
Eqs.~(\ref{eq:K57})-(\ref{eq:K59}) lead to the integral expression 
  \begin{eqnarray}
W_{np}^q(c, \pm b) =
\frac{(-1)^p}{2^{n+q+c \pm b+1}}
\int_{-1}^{1}(1-x)^{n-p}(1+x)^p\frac{d^q}{dx^q}
[(1-x)^{q+ c}(1+x)^{q \pm b}]dx
  \nonumber 
  \\
\times 
\sqrt{\frac{(2q+ c \pm b+1)(n-q)!\Gamma(q+ c \pm b+1)
\Gamma(n+q+ c \pm b+2)}{p!q!(n-p)!\Gamma(q+ c+1)
\Gamma(q\pm b+1)\Gamma(n-p+c+1)\Gamma(p \pm b+1)}}
  \label{eq:K61}
\end{eqnarray}
for the coefficient $W_{np}^q(c, \pm b)$. Equation 
(\ref{eq:K61}) can be compared with the integral representation \cite{VMK15} 
\begin{eqnarray}
(ab \alpha \beta|c \gamma)
 = \delta_{\alpha+\beta, \gamma}
\sqrt{ 
\frac{(2c+1) (J+1)! (J - 2c)! (c + \gamma)!} 
{(J - 2a)! (J - 2b)! (a-\alpha)!(a+\alpha)!(b-\beta)!(b+\beta)!(c-\gamma)!}
     }
\nonumber
\\ 
\times \frac{(-1)^{a-c+\beta}}{2^{J+1}}
       \int_{-1}^{1}(1-x)^{a-\alpha}(1+x)^{b-\beta}
       \frac{d^{c-\gamma}}{dx^{c-\gamma}}[(1-x)^{J - 2a}
                                   (1+x)^{J - 2b}]dx 
\nonumber
\end{eqnarray}
(with $J=a+b+c$)
for the Clebsch-Gordan coefficients 
$C_{a\alpha;b\beta}^{c\gamma} \equiv (ab {\alpha}{\beta}|c \gamma)$
of the compact Lie group SU(2). This yields
\begin{eqnarray}
W_{np}^q(c, \pm b)=(-1)^{n-q} \, (a_0b_0 \alpha \beta|c_0, \alpha + \beta), 
\nonumber
\\ 
a_0      = \frac{n \pm b}{2},     \ \ 
b_0      = \frac{n+c}{2},         \ \
c_0      = q + \frac{c \pm b}{2}, 
\label{eq:K63}
\\
\alpha = p - \frac{n \mp b}{2},   \ \ 
\beta  = \frac{n+c}{2} - p. 
\nonumber 
\end{eqnarray}
Since the quantum numbers in Eq.~(\ref{eq:K63})
are not necessarily integers 
or half of odd integers, the coefficients for the expansion of 
the cylindrical basis in terms of the spherical basis may be 
considered as analytical continuation, for real values of their 
arguments, of the SU(2) Clebsch-Gordan coefficients. The 
inverse of Eq.~(\ref{eq:K54}), namely
\begin{equation}
  {\Psi}_{n_r qm} 
= \sum_{p=0}^n \,
{\tilde W}_{n q}^p(c, \pm b) \, 
  {\Psi}_{n_\rho pm}
\label{eq:K64}
\end{equation}
follows from the orthonormality property of the SU(2) 
Clebsch-Gordan coefficients.
The expansion coefficients in (\ref{eq:K64}) are thus 
\begin{eqnarray}
{\tilde W}_{n q}^p (c, \pm b) = 
W_{n p}^q (c, \pm b). 
\nonumber 
\end{eqnarray}
Note that in order to compute the coefficients 
$W_{n p}^q (c, \pm b)$ through (\ref{eq:K63}), we can use the 
$_3F_2(a,b,c;d,e;1)$ representation 
\cite{VMK15} of the SU(2) Clebsch-Gordan coefficients. 

We close this section with some considerations concerning the 
limiting cases ($P = 0$, $Q \not= 0$) and 
               ($P = 0$, $Q     = 0$). It is to be observed that the passage 
from 
($P \not= 0$, $Q \not= 0$) to 
($P     = 0$, $Q \not= 0$) 
needs some caution. Indeed for $b = s^{-}$, Eq.~(\ref{eq:K63}) 
can be rewritten in terms of the quantum numbers $N$, $l$, and $n_3$ as  
\begin{eqnarray}
W_{n p}^q (c, \pm s) = (-1)^{\frac{N-l}{2}} \, 
(a_0b_0 \alpha \beta|c_0, \alpha + \beta), 
\nonumber
\\ 
a_0 = \frac{N -|m| - s \pm s}{4}, 
\nonumber
\\ 
b_0  =  \frac{N +|m| - s \mp s}{4} + \frac{\delta}{2}, 
\nonumber
\\  
\label{eq:K66}
\\
c_0  =  \frac{2l-1}{4} + \frac{\delta}{2},
\nonumber 
\\
\alpha = \frac{ 2n_3 - N + |m| - s \pm s}{4}, 
\nonumber
\\ 
\beta   =  \frac{-2n_3 + N + |m| + s \pm s}{4} 
        +  \frac{\delta}{2}. 
\nonumber 
\end{eqnarray}
By using the ordinary symmetry property \cite{VMK15}
\begin{eqnarray}
(a b   \alpha   \beta|c   \gamma) = (-1)^{a+b-c}
(a b, -\alpha, -\beta|c, -\gamma) 
\nonumber 
\end{eqnarray}
and the Regge symmetry \cite{VMK15}
\begin{eqnarray}
(ab \alpha \beta|c \gamma) =
\left( \frac{a+b       +\gamma}{2},       
       \frac{a+b       -\gamma}{2},
       \frac{a-b+\alpha-\beta }{2}, 
       \frac{a-b-\alpha+\beta }{2}| c, a-b \right)
\nonumber
\end{eqnarray}
in Eq.~(\ref{eq:K66}) with the sign $+$ and by using the ordinary 
symmetry property \cite{VMK15}
\begin{eqnarray}
(a b \alpha \beta |c \gamma) = (-1)^{a+b-c}
(b a \beta  \alpha|c \gamma)
\nonumber
\end{eqnarray}
in Eq.~(\ref{eq:K66}) with the sign $-$, we get
\begin{eqnarray}
W_{n p}^q (c, \pm s) 
\equiv
W_{N m n_3}^{l} (\delta) = (a_0 b_0 \alpha \beta|c_0, \alpha + \beta),
\nonumber
\\
     a_0  = \frac{N+|m|}{4} +\frac{\delta}{2},    \ \ 
     b_0  = \frac{N-|m|-1}{4},                    \ \
     c_0  = \frac{2l-1}{4}+\frac{\delta}{2},
\label{eq:K70}
\\
\alpha = \frac{N+|m|-2n_3}{4} + \frac{\delta}{2}, \ \
\beta  = \frac{2n_3-N+|m|-1}{4}.
\nonumber
\end{eqnarray}
As a conclusion, when $b = s^{-}$ 
we have an expansion of the type \cite{LMPS4} 
\begin{equation}
{\Psi}_{N m n_3}(\rho,\varphi,z; \delta) =
\sum_{l} 
  \, W_{N m n_3}^{l} (\delta) \, {\Psi}_{N l m}(r,\theta,\varphi; \delta),
\label{eq:K71}
\end{equation}
where the summation on $l$ goes, by steps of 2, from 
$|m|$ or $|m|+1$ to $N$ according to whether as $N-|m|$ 
is even or odd (because $N-l$ is always even). 
Equations (\ref{eq:K70})-(\ref{eq:K71}) were obtained in Ref.~\cite{LMPS4} for 
the ring-shape oscillator system. Finally, the case
$P = Q = 0$ can be easily deduced from (\ref{eq:K70})-(\ref{eq:K71}) by taking
$\delta = 0$:
we thus recover the result obtained in Refs.~\cite{PT17,PST20}
for the isotropic harmonic oscillator in three dimensions.
Note that in the case $P = Q = 0$, the expansion
coefficients in Eq.~(\ref{eq:K71}) become Clebsch-Gordan coefficients
for the noncompact Lie group SU(1,1) (cf., Ref.~\cite{KPSbis36}). 

\vspace{0.3cm} 
{\bf Prolate and Oblate Spheroidal Bases}
\vspace{0.2cm}

{\bf SEPARATION IN PROLATE SPHEROIDAL COORDINATES}
\vspace{0.1cm}

The prolate spheroidal coordinates $(\xi, \eta, \varphi)$ are
defined via
\begin{eqnarray}
x &=& \frac{R}{2}\sqrt{({\xi}^2-1)(1-{\eta}^2)}\cos\varphi,
\nonumber
\\
y &=& \frac{R}{2}\sqrt{({\xi}^2-1)(1-{\eta}^2)}\sin\varphi, 
\nonumber
\\
z &=& \frac{R}{2}{\xi}{\eta},
\nonumber
\end{eqnarray}
(with $1\leq \xi <\infty$, 
      $-1\leq \eta \leq1$, 
and   $0\leq \varphi<2\pi$), 
where $R$ is the interfocus distance. 
As is well-known \cite{KPS28}, in the limits where
$R\to0$ and $R\to\infty$, the prolate spheroidal coordinates
reduce to the spherical coordinates and the cylindrical
coordinates, respectively. In prolate spheroidal coordinates, 
the potential $V$ reads
\begin{equation}
V=\frac{ {\Omega}^2 R^2 }{8} ({\xi}^2+{\eta}^2-1)+
\frac{2}{R^2}\left[
 \frac{P}{{\xi}^2{\eta}^2} +
 \frac{Q}{({\xi}^2-1)(1-{\eta}^2)}
\right].
\label{eq:K73}
\end{equation}
By looking for a solution $\Psi$ of Eq.~(\ref{eq:K2}), 
with the potential (\ref{eq:K73}), in the form
\begin{equation}
\Psi(\xi,\eta,\varphi)={\psi}_1(\xi){\psi}_2(\eta)
\frac{{\rm e}^{im\varphi}}{\sqrt {2\pi}}, 
\label{eq:K74}
\end{equation}
with $m \in {\bf Z}$, we obtain the two ordinary differential equations
\begin{equation}
\Biggl[ d_{\xi}  ({\xi}^2-1)  d_{\xi}  -
\frac{Q +m^2}{{\xi}^2-1}+\frac{ER^2}{2}{\xi}^2
\nonumber
\\
-
\frac{{\Omega}^2R^4}{16}{\xi}^2({\xi}^2-1)+\frac{P}{{\xi}^2}
\Biggr] {\psi}_1 = + \lambda(R) {\psi}_1,
\label{eq:K75}
\end{equation} 
\begin{equation}
\Biggl[ d_{\eta} (1-{\eta}^2) d_{\eta} -
\frac{Q+m^2}{1-{\eta}^2}-\frac{ER^2}{2}{\eta}^2
\nonumber
\\
-
\frac{{\Omega}^2R^4}{16}{\eta}^2(1-{\eta}^2)-\frac{P}{{\eta}^2}
\Biggr] {\psi}_2 = - \lambda(R) {\psi}_2,
\label{eq:K76}
\end{equation}
where $\lambda(R)$ is a separation constant in prolate
spheroidal coordinates. The combination of 
Eqs.~(\ref{eq:K75}) 
and  (\ref{eq:K76}), leads to the operator
\begin{eqnarray}
\Lambda =
&-&
\frac{1}{{\xi}^2-{\eta}^2} \left[
{\eta}^2 \partial_{\xi} ({\xi}^2-1) \partial_{\xi} + 
 {\xi}^2 \partial_{\eta}(1-{\eta}^2)\partial_{\eta} \right] 
\nonumber\\
&+&
\frac{{\xi}^2+{\eta}^2-1}{({\xi}^2-1)(1-{\eta}^2)}
(Q - \partial_{\varphi\varphi})
\nonumber\\ 
&+& \frac{ {\Omega}^2 R^4 }{16} {\xi}^2 {\eta}^2 + 
P \frac{{\xi}^2+{\eta}^2}{{\xi}^2{\eta}^2} 
\nonumber
\end{eqnarray}
after eliminating the energy $E$. The eigenvalues of the operator $\Lambda$
are $\lambda(R)$ while its eigenfunctions are given by (\ref{eq:K74}). 
The significance of the (self-adjoint) operator $\Lambda$ is to be found in 
the connecting formula 
\begin{equation}
\Lambda=M+\frac{R^2}{4}N,
\label{eq:K78}
\end{equation}
where $M$ and $N$ are the constants of motion (\ref{eq:K34}) 
                                          and (\ref{eq:K52}). 
The operator $\Lambda$ is of pivotal importance for the derivation
of the interbasis expansion coefficients from the spherical
basis or the cylindrical basis to the prolate spheroidal basis.
In particular, it allows us to derive the latter coefficients without 
knowing the wavefunctions in prolate spheroidal basis. (In this
respect, credit should be put on the work by Coulson and Joseph 
\cite{CJ34} who considered an operator similar to $\Lambda$ for 
the hydrogen atom.) Therefore, we shall not derive the prolate 
spheroidal wavefunctions ${\psi}_1$ and 
                         ${\psi}_2$ which could be 
obtained by solving Eqs.~(\ref{eq:K75}) 
                     and (\ref{eq:K76}). It is more economical 
to proceed in the following way that presents the advantage 
of giving, at the same time, the global wavefunction
$\Psi(\xi, \eta, \varphi) \equiv 
 \Psi(\xi, \eta, \varphi; R, c, \pm b)$ and the 
interbasis expansion coefficients.
\vspace{0.2cm}

{\bf INTERBASIS EXPANSIONS FOR\par 
     THE PROLATE SPHEROIDAL WAVEFUNCTIONS}
\vspace{0.1cm}

The three constants of motion 
$M$, $N$, and $\Lambda$, which occur in Eq.~(\ref{eq:K78}),
can be seen to satisfy the following eigenequations 
\begin{eqnarray}
  M{\Psi}_{n_rqm} &=& A_q(c, \pm b)
   {\Psi}_{n_rqm}, 
\label{eq:K79}
\\
  N{\Psi}_{n_\rho pm} &=& 2 E_z(p, \pm b)
   {\Psi}_{n_\rho pm},
\label{eq:K80}
\end{eqnarray}
and
\begin{equation}
         \Lambda{\Psi}_{nkm} =
  {\lambda}_k(R){\Psi}_{nkm} 
\label{eq:K81}
\end{equation}
for the spherical, cylindrical, and prolate spheroidal bases,
respectively. [In Eq.~(\ref{eq:K81}), the index $k$ labels the
eigenvalues of the operator $\Lambda$ and varies in the
range $0 \leq k \leq n$.] The spherical, cylindrical, and 
prolate spheroidal bases are indeed eigenbases for the three 
sets of commuting operators $\{H,L_z,M\}$, 
                            $\{H,L_z,N\}$, and 
                            $\{H,L_z,\Lambda\}$, 
respectively. We are now in a position to deal with the interbasis expansions
\begin{eqnarray}
  {\Psi}_{nkm} &=& \sum_{p=0}^n{U}_{nk}^p(R; c, \pm b) 
  {\Psi}_{n_\rho pm} 
\label{eq:K82} \\
  {\Psi}_{nkm} &=& \sum_{q=0}^n{T}_{nk}^q(R; c, \pm b)
  {\Psi}_{n_r qm} 
\label{eq:K83}
\end{eqnarray}
for the prolate spheroidal basis in terms of the cylindrical
and spherical bases.

First, we consider Eq.~(\ref{eq:K82}). Let the operator $\Lambda$ act 
on both sides of       (\ref{eq:K82}). Then, by using 
                  Eqs.~(\ref{eq:K78}), 
                       (\ref{eq:K80}), and      
                       (\ref{eq:K81}) 
along with the orthonormality property 
of the cylindrical basis, we find that
\begin{eqnarray}
\frac {1}{2} \biggl[{\lambda}_k(R)-\frac{R^2}{2} E_z(p, \pm b) 
\biggr]U_{nk}^p(R; c, \pm b) 
\nonumber
\\
=
\sum_{p'=0}^nU_{nk}^{p'}(R; c, \pm b)
M_{pp'}^{(\pm)},
\label{eq:K84}
\end{eqnarray}
where
\begin{equation}
M_{pp'}^{(\pm)} = \int_0^\infty\!\int_0^{2\pi}\!\int_0^\infty
  {\Psi}_{n_\rho pm}^{*} 
 M{\Psi}_{n_\rho p'm} 
\rho {d \rho}{d\varphi}dz.
\label{eq:K85}
\end{equation}
The calculation of the matrix element $M_{pp'}^{(\pm)}$ can be
done by expanding the cylindrical wavefunctions in (\ref{eq:K85}) 
in terms of spherical wavefunctions [see   Eq.~(\ref{eq:K54})]
and by making use of the eigenvalue 
equation for $M$                    [see   Eq.~(\ref{eq:K79})]. This leads to
\begin{equation}
M_{pp'}^{(\pm)} = \frac{1}{2}\sum_{q=0}^n
        A_q(c, \pm b)
{W}_{np }^q(c, \pm b)
{W}_{np'}^q(c, \pm b).
\label{eq:K86}
\end{equation}
To calculate the sum in Eq.~(\ref{eq:K86}), we need
some recursion relation for the coefficient ${W}_{np}^q(c, \pm b)$
involving $p-1$, $p$, and $p+1$. Owing to Eq.~(\ref{eq:K63}), this 
amounts to use the following recursion relations \cite{KCS37}:
\begin{eqnarray}
[-a(a+1) -b(b+1) + c(c+1) - 2 \alpha \beta]
(ab \alpha \beta|c \gamma)
\nonumber \\ 
=
\sqrt{(a+\alpha)(a-\alpha+1)(b-\beta)(b+\beta+1)}
\nonumber
\\
\times (a, b, \alpha-1, \beta+1|c \gamma)
\nonumber \\ 
+
\sqrt{(a-\alpha)(a+\alpha+1)(b+\beta)(b-\beta+1)}
\nonumber
\\
\times (a, b, \alpha+1, \beta-1|c \gamma).
\label{eq:K87}
\end{eqnarray}
                    Then, by introducing Eq.~(\ref{eq:K87}) into 
                                         Eq.~(\ref{eq:K86}) 
and by using the orthonormality condition
\begin{equation}
\sum_{c,\gamma}
(ab \alpha  \beta |c \gamma)
(ab \alpha' \beta'|c \gamma)
={\delta}_{\alpha{\alpha}^{'}}
{\delta}_{\beta{\beta}^{'}},
\label{eq:K88}
\end{equation}
we find that $M_{pp'}^{(\pm)}$ is given by
\begin{eqnarray}
M_{pp'}^{(\pm)} = 
2\bigl[ p (p \pm b) (n-p+1) (n+c-p+1) \bigr]^{s} {\delta}_{p',p-1}
\nonumber
\\ 
+ [s \left(c \mp b +  s\right)
     \left(c \mp b + 3s\right) + 2(p+1)(n-p) + 
  2(p \pm b)(n+c-p+1)] {\delta}_{p'p}
\nonumber
\\ 
+ 2\bigl[ (p+1) (p + 1 \pm b) (n-p) (n+c-p) \bigr]^{s} {\delta}_{p',p+1}.
\label{eq:K89} 
\end{eqnarray}
Now by introducing (\ref{eq:K89}) into 
                   (\ref{eq:K84}), we get the following three-term
recursion relation 
\begin{eqnarray}
\bigl[(p+1)(n-p)+(p \pm b)(n+c-p+1)
\nonumber 
\\ 
+ \frac{1}{4} (c \mp b+ s)
              (c \mp b+3s)
+ \frac {R^2} {8} E_z(p, \pm b) - \frac {1} {4} {\lambda}_k(R)
\bigr] 
  U_{nk}^p 
\nonumber \\ 
+ \bigl[ (p+1) (p + 1 \pm b) (n-p) (n+c-p) \bigr]^{s}
  U_{nk}^{p+1}
\nonumber \\ 
+ \bigl[ p (p \pm b) (n-p+1) (n+c-p+1) \bigr]^{s}
  U_{nk}^{p-1}              = 0 
\label{eq:K90}
\end{eqnarray}
for the expansion coefficients $U_{nk}^{q} \equiv U_{nk}^q (R; c, \pm b)$. 
The recursion relation (\ref{eq:K90}) provides us with a system
of $n+1$ linear homogeneous equations which can be
solved by taking into account the normalization condition
\begin{eqnarray}
\sum_{p=0}^n|U_{nk}^p(R; c, \pm b)|^2=1.
\nonumber
\end{eqnarray}
The eigenvalues ${\lambda}_k(R)$ of the operator $\Lambda$
then follow from the vanishing of the determinant for the
latter system.

Second, let us concentrate on the expansion (\ref{eq:K83}) of 
the prolate spheroidal basis in terms of the spherical basis. By 
employing a technique similar to the one used for deriving
Eq.~(\ref{eq:K84}), we get
\begin{eqnarray}
\left[ {\lambda}_k(R) - A_q(c, \pm b) \right] T_{nk}^q(R; c, \pm b) 
\nonumber
\\
= \frac {R^2} {2}
\sum_{q'=0}^nT_{nk}^{q'}(R; c, \pm b)\,
N_{qq'}^{(\pm)},
\label{eq:K92}
\end{eqnarray}
where
\begin{eqnarray}
N_{qq'}^{(\pm)} = \int_0^\infty              \!
                  \int_0^{ \frac {\pi} {2} } \!
                  \int_0^{2\pi}
  {\Psi}_{n_r qm}^{*}
N
  {\Psi}_{n_r q'm} 
r^2 \sin\theta{dr}{d \theta}{d\varphi}.
\nonumber
\end{eqnarray}
The matrix elements $N_{qq'}^{(\pm)}$ can be calculated in the
same way as $M_{pp'}^{(\pm)}$ except that we must use the 
relation \cite{VMK15} 
\begin{eqnarray} 
(ab \alpha \beta | c \gamma)
= -\Biggl[\frac{c^2(2c+1)
(2c-1)}{ (c^2 - \gamma^2) (-a+b+c) (a-b+c) (a+b-c+1)
(a+b+c+1)}\Biggr]^{s}
\nonumber
\\
\times
\Biggl\{\Biggl[\frac{(c-\gamma-1)
(c+\gamma-1)(-a+b+c-1)(a-b+c-1)(a+b-c+2)(a+b+c)}
{(c-1)^2(2c-3)(2c-1)}\Biggr]^{s}
\nonumber
\\
\times 
(ab \alpha \beta|c-2, \gamma)
-\frac{(\alpha-\beta)c(c-1)-{\gamma}a(a+1)+
{\gamma}b(b+1)}{c(c-1)}
(ab \alpha \beta|c-1, \gamma)
\Biggr\}     
\nonumber
\end{eqnarray}
and the orthonormality condition
\begin{eqnarray}
\sum_{\alpha,\beta}
(ab \alpha \beta|c \gamma) (ab \alpha \beta|c' \gamma')
=
{\delta}_{c'c}{\delta}_{{\gamma}' \gamma},
\nonumber
\end{eqnarray}
instead of Eqs.~(\ref{eq:K87}) and 
                (\ref{eq:K88}). This produces the matrix element
\begin{eqnarray}
N_{qq'}^{(\pm)} = E_n(c, \pm b) \, 
\frac{2q(q+1)+(c \pm b)(2q \pm b+1)}
{(2q+ c \pm b)(2q+ c \pm b+2)}
\delta_{q'q} 
\nonumber
\\ 
- 2 \Omega \bigl[ A_{n}^{q+1}(c, \pm b) {\delta}_{q',q+1}+
                  A_{n}^q    (c, \pm b) {\delta}_{q',q-1} \bigr],
\label{eq:K96}
\end{eqnarray}
where
\begin{eqnarray}
A_{n}^q(c, \pm b) = \Biggl[\frac{q (n-q+1)(q+c \pm b)
(q \pm b)(q+ c)(n+q+c \pm b+1)}
{(2q+ c \pm b)^2(2q + c \pm b - 1)
                    (2q + c \pm b + 1)}\Biggr]^{s}.
\nonumber
\end{eqnarray}
Finally, the introduction of (\ref{eq:K96}) into 
                             (\ref{eq:K92}) leads
to the three-term recursion relation
\begin{eqnarray}
\Biggl[ {\lambda}_k(R) - A_q(c, \pm b) 
- \frac {R^2} {2} E_n(c, \pm b)
\nonumber
\\
\times 
\frac{2q(q+1)+(c \pm b)(2q    \pm b+1)}
        {(2q + c \pm b)(2q+ c \pm b+2)} \Biggr]
  T_{nk}^q
\nonumber \\ 
+ \Omega R^2 \Bigl[
A_{n}^{q+1}(c,\pm b)
  T_{nk}^{q+1}             +
A_{n}^ q   (c,\pm b)
  T_{nk}^{q-1} 
\Bigr] = 0
\label{eq:K98}
\end{eqnarray}
for the expansion coefficients
$T_{nk}^{p} \equiv T_{nk}^{p}(R; c, \pm b)$. This 
relation can be iterated by taking 
into account the normalization condition
\begin{eqnarray}
\sum_{q=0}^n|T_{nk}^q(R; c,\pm b)|^2=1.
\nonumber
\end{eqnarray}
Here again, the eigenvalues $\lambda_k(R)$ may be
obtained from the vanishing of the determinant of 
a system of $n+1$ linear homogeneous equations.

\vfill\eject     
{\bf LIMITING CASES}
\vspace{0.1cm}

Putting $b = s^{-}$, i.e., $P=0^{-}$, 
in the matrix element (\ref{eq:K96}) 
with $Q \not= 0$ 
and by using (\ref{eq:K26}), we have
\begin{eqnarray}
N_{qq'}^{(\pm)} = E_N(\delta) 
\frac{2 A_l(\delta) - 2(|m|+ \delta)^2 -1} {(2l+2\delta-1)(2l+2\delta+3)}
\delta_{l'l} 
\nonumber
\\
-
2 \Omega \bigl[ A_{N}^{l+2}(\delta) {\delta}_{l',l+2} +
                A_{N}^ l   (\delta) {\delta}_{l',l-2} \bigr],
\nonumber
\end{eqnarray}
where
\begin{eqnarray}
A_{N}^l(\delta)                               = 
\Biggl[\frac{ l_-
             (l_- - 1)(l_+ + 2\delta)
                      (l_+ + 2\delta - 1)
(N-l+2)(N+l+2\delta+1)}
{4(2l+2\delta-1)^2(2l+2\delta-3)(2l+2\delta+1)}\Biggr]^{s}
\nonumber 
\end{eqnarray}
(with $l_{\pm} = l \pm |m|$) 
and finally we get the following three-term recursion relation 
\begin{eqnarray}
\Biggl[{\lambda}_k(R) - A_l(\delta) - \frac{R^2}{2} E_N(\delta) 
                \frac{2 A_l(\delta) - 2(|m|+\delta)^2-1}
{(2l+2\delta-1)(2l+2\delta+3)}\Biggr]
T_{Nk}^l(R;\delta)
\nonumber
\\ 
+ \Omega R^2\Bigl[
A_{N}^{l+2}(\delta)T_{Nk}^{l+2}(R;\delta)+
A_{N}^ l   (\delta)T_{Nk}^{l-2}(R;\delta)\Bigr]=0 
\nonumber
\end{eqnarray}
for $T_{Nk}^l(R;\delta) \equiv T_{Nk}^l(R; c, \pm s)$. By 
analogy it is easy to obtain a three-term recursion relation for 
the interbasis expansion coefficients $U_{Nk}^{n_3}(R;\delta) \equiv
U_{Nk}^{n_3}(R; c, \pm s)$. We get
\begin{eqnarray}
\bigl[(2n_3+1)(N-n_3+\delta+1)+(|m|+\delta)^2-1
+\frac{\Omega R^2}{4}(2n_3+1)-{\lambda}_k(R)
\bigr]U_{Nk}^{n_3}(R;\delta)
\nonumber \\ 
+ \bigl[(n_3+1)(n_3+2)(N-|m|-n_3)(N+|m|-n_3+2\delta)\bigr]^{s}
U_{Nk}^{n_3+2}(R;\delta)
\nonumber \\ 
+ \bigl[n_3(n_3-1)(N-|m|-n_3+2)(N+|m|-n_3+2\delta+2)\bigr]^{s}
U_{nk}^{n_3-2}(R;\delta)=0.
\nonumber
\end{eqnarray}
Consequently, when $b = s^{-}$ we have the 
expansions [cf., Eqs.~(\ref{eq:K82}) and 
                      (\ref{eq:K83})] 
\begin{eqnarray}
  {\Psi}_{Nkm}                              &=&
\sum_{n_3}^N {U}_{Nk}^{n_3} (R; \delta)
  {\Psi}_{Nmn_3}                        ,
\nonumber \\
  {\Psi}_{Nkm}                              &=&
\sum_{l}^N   {T}_{Nk}^l     (R; \delta)
  {\Psi}_{Nlm}                          ,
\nonumber
\end{eqnarray}
for the ring-shape oscillator. The summations on $l$ and
$n_3$ go, by steps of 2, from $|m|$ or $|m|+1$ to $N$ and
from 0 or 1 to $N-|m|$ according to whether as $N-|m|$ is
even or odd (because $N-l$ and $N-|m|-n_3$ are always even).

The next limiting case $\delta = 0$, i.e., $Q = 0$, is trivial
and the corresponding results for the isotropic harmonic oscillator 
agree with the ones obtained in Ref.~\cite{MPST18}.

Finally, it should be noted that the following two limits
\begin{eqnarray}
\lim_{R\to 0}U_{nk}^p(R; c, \pm b) &=&
       {\tilde W}_{nk}^p(c, \pm b), 
\nonumber 
\\
\lim_{R\to\infty}T_{nk}^q(R; c, \pm b) &=&
                    W_{nk}^q(c, \pm b)
\nonumber
\end{eqnarray}
furnish a useful means for checking the calculations
presented in the fourth and fifth sections. 
\vspace{0.2cm}

{\bf SEPARATION AND INTERBASIS EXPANSIONS FOR\par 
     THE OBLATE SPHEROIDAL WAVEFUNCTIONS}
\vspace{0.1cm}

The oblate spheroidal coordinates 
$(\overline\xi, \overline\eta, \varphi)$ are defined by 
\begin{eqnarray}
x &=& \frac{\overline R}{2}
\sqrt{({\overline\xi}^2+1)(1-{\overline\eta}^2)}\cos\varphi,
\nonumber
\\ 
y &=& \frac{\overline R}{2}
\sqrt{({\overline\xi}^2+1)(1-{\overline\eta}^2)}\sin\varphi,
\nonumber
\\ 
z &=& \frac{\overline R}{2} \, {\overline\xi} \, {\overline\eta},
\nonumber
\end{eqnarray}
(with  $0\leq \overline\xi <\infty$, 
       $-1\leq \overline\eta \leq1$, 
and    $0\leq \varphi<2\pi$), 
where $\overline R$ is the interfocus distance in the oblate spheroidal
coordinate system. As in the prolate system, in the limits 
$\overline R\to 0$ and $\overline R \to \infty$, the oblate spheroidal 
coordinates give the spherical and cylindrical
coordinates, respectively \cite{KPS28,MPST18}. 

The potential $V$, the Schr\"odinger equation, the oblate spheroidal
constant of motion $\overline \Lambda$, 
and the interbasis expansion coefficients
for the oblate spheroidal coordinates can be obtained from the corresponding 
expressions for the prolate spheroidal coordinates by means of the trick: 
$ \xi \rightarrow - i \overline \xi $ and 
$   R \rightarrow   i \overline   R $. 

\vspace{0.3cm}
{\bf Spheroidal Corrections for the Spherical and Cylindrical Bases}
\vspace{0.1cm}

As we have already mentioned, the spheroidal system of coordinates
is one of the most general one-parameter systems of coordinates
which contains spherical and cylindrical coordinates as some limiting
cases. 
Accordingly, the prolate spheroidal basis of the generalized oscillator as
$R \rightarrow 0$ and $R\rightarrow \infty$ degenerates into the
spherical and cylindrical bases that can be treated as zeroth order 
approximations in some perturbation series.
The three-term recursion relations for the expansion
coefficients of the prolate spheroidal basis in the cylindrical
and spherical bases, which have been obtained in the fifth section, 
may serve as a basis for constructing an algebraic
perturbation theory, respectively, at large ($R \gg 1$) and small  
                                            ($R \ll 1$) values of
the interfocus distance $R$. Thus it is possible to derive prolate 
spheroidal corrections for the spherical and cylindrical bases.
\vspace{0.2cm}

{\bf THE CASE $R \ll 1$}
\vspace{0.1cm}

Let us rewrite the three-term recursion relation (\ref{eq:K98}) in the
following form
\begin{eqnarray}
[ {\lambda}_k(R) - A_q(c, \pm b) - \Omega R^2 B_{n}^{q}(c, \pm b) ]
  T_{nk}^q 
\nonumber\\ 
+ \Omega R^2 \Bigl[
A_{n}^{q+1}(c,\pm b)
  T_{nk}^{q+1}             +
A_{n}^q(c,\pm b)
  T_{nk}^{q-1}             \Bigr] = 0, 
\label{eq:K109}
\end{eqnarray}
where
\begin{eqnarray}
B_{n}^{q}(c,\pm b)=
\frac{1}{2} ( 2n + c \pm b + 2 )
\nonumber
\\
\times 
\frac{2q(q+1)+(c \pm b)(2q \pm b+1)}
{(2q+ c \pm b)(2q+ c \pm b+2)}.
\nonumber
\end{eqnarray}
The zeroth order approximation for the separation constant ${\lambda}_k(R)$
and the coefficients $T_{nk}^{p}(R; c, \pm b)$ can immediately be
derived from the recursion relation (\ref{eq:K109}). Indeed, from 
                                Eq.~(\ref{eq:K109}), we obtain 
\begin{eqnarray}
\lim_{R\to 0}{\lambda}_k(R) = A_k(c, \pm b), 
\nonumber
\\ 
\lim_{R\to 0}T_{nk}^q(R; c, \pm b) = {\delta}_{kq},
\nonumber
\end{eqnarray}
so that, for the wavefunction, we have
\begin{eqnarray}
\lim_{R \to 0}{\Psi}_{nkm}(\xi,\eta,\varphi; R,c, \pm b)
= {\Psi}_{n_{r}km}(r,\theta,\varphi,z; c, \pm b).
\nonumber
\end{eqnarray}
As is seen from these limiting relations, the quantum
number $k$, labeling the spheroi\-dal separation constant and 
being (according to the oscillation theorem \cite{KPS28}) 
the number of zeros of the prolate angular spheroidal function
$\psi_2(\eta)$ in the interval $-1\leq \eta \leq 1$, turns into a
spherical quantum number determining the number of zeros of the
angular function (\ref{eq:K11}). It is clear that this fact is a 
consequence of the independence of the number of zeros of the 
wavefunction on $R$.

In order to calculate higher order corrections, 
we represent the interbasis coefficients $T_{nk}^{q} (R; c, \pm b)$ and
the spheroidal separation constant ${\lambda}_{k} (R)$ as expansions
in powers of $\Omega R^2$:
\begin{equation}
T_{nk}^q(R; c, \pm b) = {\delta}_{kq}
+ \sum_{j=1}^{\infty}{T}_{kq}^{(j)}~(\Omega R^2)^j,
\label{eq:K114}
\end{equation}
\begin{equation}
{\lambda}_{k}(R) = A_k(c, \pm b) 
+ \sum_{j=1}^{\infty}{\lambda}_k^{(j)}~(\Omega R^2)^j.
\label{eq:K115}
\end{equation}
Substituting      Eqs.~(\ref{eq:K114}) and 
                       (\ref{eq:K115}) into the three-term recursion
              relation (\ref{eq:K109}) and equating the coefficients 
with the same power of $R$, we arrive at the equation
for the coefficients $T_{kq}^{(j)}$ and  ${\lambda}_{k}^{(j)}$
\begin{eqnarray}
4(k-q)(k+q+c \pm b +1) T_{kq}^{(j)} =
\nonumber 
\\ 
- A_n^{q+1} (c,\pm b) T_{k,q+1}^{(j-1)}
+ B_n^ q    (c,\pm b) T_{kq   }^{(j-1)}
\nonumber
\\
- A_n^q (c,\pm b) T_{k,q-1}^{(j-1)}
- \sum_{t=0}^{j-1}{\lambda}_{k}^{(j-t)} T_{kq}^{(t)}. 
\label{eq:K116}
\end{eqnarray}
Equation (\ref{eq:K116}) with the initial condition
$T_{kq}^{(0)} = {\delta}_{kq}$
and the condition
$T_{qq}^{(j)} = {\delta}_{j0}$
arising in the standard perturbation theory \cite{LL38} allow us
to derive a formula expressing ${\lambda}_k^{(j)}$ for $j \geq 1$
through the coefficients $T_{kk}^{(j-1)}$ and $T_{k,k \pm 1}^{(j-1)}$:
\begin{equation}
{\lambda}_{k}^{(j)} = - A_n^{k+1} (c,\pm b) T_{k,k+1}^{(j-1)}
\nonumber
\\
+ B_n^{k} (c,\pm b) T_{k,k}^{(j-1)}
\nonumber
\\
- A_n^{k} (c,\pm b) T_{k,k-1}^{(j-1)}.
\label{eq:K117}
\end{equation}
This gives a possibility to determine, step by step, the
coefficients ${\lambda}_k^{(j)}$ and $T_{kq}^{(j)}$ in 
Eqs.~(\ref{eq:K114}) and 
     (\ref{eq:K115}). As an example, let us 
write down the first and second order corrections in
(\ref{eq:K115}) for 
${\lambda}_k (R)$ and the 
first order correction in (\ref{eq:K114}) for 
$T_{nk}^{q}(R; c, \pm b)$. 
It follows from Eq.~(\ref{eq:K117}) that
\begin{eqnarray}
{\lambda}_{k}^{(1)} =   B_n^{k }  (c,\pm b), 
\nonumber
\\ 
{\lambda}_{k}^{(2)} = - A_n^{k+1} (c,\pm b) T_{k,k+1}^{(1)}
                      - A_n^{k  } (c,\pm b) T_{k,k-1}^{(1)}
\nonumber
\end{eqnarray}
and Eq.~(\ref{eq:K116}) for $j = 1$ results in
\begin{eqnarray}
T_{kq}^{(1)} = - \frac{A_n^{k}(c,\pm b)}
{4(2k+c\pm b)}
{\delta}_{q,k-1}
+ \frac{A_n^{k+1}(c,\pm b)}{4(2k+c\pm b+2)}
{\delta}_{q,k+1}.
\label{eq:K120}
\end{eqnarray}
Thus, 
for the spheroidal separation constant, with an accuracy up to
the term $(\Omega R^2)^2$, we get 
\begin{eqnarray}
{\lambda}_{k}(R) = A_k(c, \pm b) + {\Omega}R^2B_n^{k}(c, \pm b)
\nonumber
\\
+\frac{{\Omega}^2R^4}{4}\Biggl [ 
 \frac{{A_n^{k  }(c,\pm b)}^2}{2k+c\pm b  }
-\frac{{A_n^{k+1}(c,\pm b)}^2}{2k+c\pm b+2}\Biggr ].
\nonumber
\end{eqnarray}
Introducing  (\ref{eq:K120}) into 
             (\ref{eq:K114}) and then using 
             (\ref{eq:K83}) for the expansion
of the prolate 
spheroidal basis over the spherical one, we get the following 
approximate formula 
\begin{eqnarray}
{\Psi}_{nkm}(\xi,\eta,\varphi; R, c, \pm b)=
{\Psi}_{nkm}(r,\theta,\varphi;    c, \pm b) 
\nonumber 
\\
- \frac{{\Omega}^2R^4}{4} \Biggl[
 \frac{A_n^{k  }(c,\pm b)}{2k+c\pm b}
{\Psi}_{n,k-1,m}(r,\theta,\varphi; c, \pm b)
\nonumber
\\
-\frac{A_n^{k+1}(c,\pm b)}{2k+c\pm b+2}
{\Psi}_{n,k+1,m}(r,\theta,\varphi; c, \pm b)\Biggr]. 
\nonumber
\end{eqnarray}
\vspace{0.1cm}

{\bf THE CASE $R \gg 1$}

Now let us consider the case $R \gg 1$. The three-term recursion
relation (\ref{eq:K90}) can be written as
\begin{eqnarray}
\bigl[D_n^{p}(c,\pm b)
+\frac{R^2}{8} E_z(p, \pm b) 
-\frac{{\lambda}_k(R)}{4}
\bigr]
  U_{nk}^p 
\nonumber\\ 
+ \bigl[C_n^{p+1}(c, \pm b)
  U_{nk}^{p+1} 
+ C_n^p(c, \pm b)
  U_{nk}^{p-1}
\bigr]=0,
\label{eq:K123}
\end{eqnarray}
where
\begin{eqnarray}
C_n^p(c,\pm b) = 
\bigl[p(p \pm b)(n-p+1)(n+c-p+1)\bigr]^{s}, 
\nonumber
\\
D_n^p(c,\pm b) = (p+1)(n-p)
\nonumber
\\
+(p \pm b)(n+c-p+1) 
\nonumber
\\
+ \frac{1}{4} ( c \mp b+ s )
              ( c \mp b+3s ).
\nonumber
\end{eqnarray}
It follows from Eq.~(\ref{eq:K123}) that 
\begin{eqnarray}
\lim_{R\to \infty} \frac{{\lambda}_k(R)}{ R^2 }=\frac{1}{2}E_z(k, \pm b), 
\nonumber
\\ 
\lim_{R\to \infty} U_{nk}^p(R; c, \pm b) = {\delta}_{kp}.
\nonumber
\end{eqnarray}
For $R \gg 1$, the interbasis expansion coefficients and the spheroidal
separation constant are developed in negative powers of $\Omega R^2$:
\begin{equation}
U_{nk}^p(R; c, \pm b) = {\delta}_{kp}
+ \sum_{j=1}^{\infty}{U}_{kp}^{(j)}(\Omega R^2)^{-j},
\label{eq:K128}
\end{equation}
\begin{equation}
\frac{{\lambda}_{k}(R)}{\Omega R^2}
=\frac{1}{2 \Omega} E_z(k, \pm b) 
+ \sum_{j=1}^{\infty}{\lambda}_k^{(j)}(\Omega R^2)^{-j}.
\label{eq:K129}
\end{equation}
Substituting Eqs.~(\ref{eq:K128}) and 
                  (\ref{eq:K129}) into 
              Eq.~(\ref{eq:K123}), we get 
\begin{eqnarray}
\frac{1}{4}(p-k)U_{kp}^{(j)} + 
C_n^{p+1}(c,\pm b) U_{k,p+1}^{(j-1)} +
D_n^ p   (c,\pm b) U_{k p  }^{(j-1)}
\nonumber
\\ 
+ C_n^{p}(c,\pm b) U_{k,p-1}^{(j-1)}
-\frac{1}{4} \sum_{t=1}^{j-1}{\lambda}_k^{(j-t)}
U_{kp}^{(t)} = 0.
\label{eq:K130}
\end{eqnarray}
Using the conditions
$U_{kp}^{(0)}={\delta}_{kp}$ and 
$U_{pp}^{(j)}={\delta}_{j0}$, 
one easily obtain
\begin{equation}
\frac{1}{4}{\lambda}_k^{(j)} =
C_n^{p+1}(c,\pm b) U_{k,p+1}^{(j-1)}
\nonumber
\\
+
D_n^ p   (c,\pm b) U_{k p  }^{(j-1)}+ 
C_n^{p}  (c,\pm b) U_{k,p-1}^{(j-1)}.
\label{eq:K132}
\end{equation}
Equations (\ref{eq:K130}) 
      and (\ref{eq:K132}) completely solve 
the problem of determining the expansion
coefficients ${\lambda}_k^{(j)}$ and $U_{kp}^{(j)}$. For 
instance, we have the approximate formulae 
\begin{eqnarray}
\frac{{\lambda}_{k}(R)}{\Omega R^2}
= \frac{1}{2 \Omega} E_z(k,  \pm b) +
\frac{4}{\Omega R^2} D_n^k(c,\pm b)
\nonumber
\\
+ \frac{16}{(\Omega R^2)^2}
\left[  {C_n^{k}  (c,\pm b)}^2
      - {C_n^{k+1}(c,\pm b)}^2
\right],
\nonumber 
\\ 
{\Psi}_{nkm}(\xi,\eta,\varphi; R, c, \pm b)=
{\Psi}_{nkm}(\rho,\varphi,z; c, \pm b)     
\nonumber 
\\
+ \frac{4}{\Omega R^2} \Biggl[ 
C_n^{k}(c,\pm b)
{\Psi}_{n,k-1,m}(\rho,\varphi,z; c, \pm b)
\nonumber
\\
-C_n^{k+1}(c,\pm b)
{\Psi}_{n,k+1,m}(\rho,\varphi,z; c, \pm b)
\Biggr].
\nonumber 
\end{eqnarray}

\vspace{0.3cm}
{\bf ACKNOWLEDGMENTS}
\vspace{0.1cm}

One of the authors (G.S.P.) is grateful to the 
{\it Institut de Physique Nucl\'eaire de Lyon} for the kind 
hospitality extended to him during his stay in Lyon-Villeurbanne. 
A preliminary version of this work was discussed at the International Workshop 
``Finite Dimensional Integrable Systems'' (Dubna, Russia, July 1994). 
In this respect, the authors thank  A.N.~Sissakian, V.M.~Ter-Antonyan,
and P.~Winternitz for interesting discussions. 

\vspace{0.3cm}
{\bf Appendix: Bi-Orthogonality of the Radial Wavefunctions}
\vspace{0.1cm}

Besides the orthonormality relation (\ref{eq:K17}) in the quantum 
numbers
$n_r$ for the function $R_{n_r q}$, we also have an orthogonality
relation in the quantum numbers $q$, viz.,
\begin{equation}
J_{qq'}^{(\pm)} = \int_{0}^{\infty}
  R_{n_r' q'}
  R_{n_r  q }             dr 
= \frac{\Omega}{2q+c \pm b+1}
\delta_{q'q}, 
\label{eq:K135}
\end{equation}
for a given value $n_r' + q' = n_r + q$ 
of the principal quantum number $n$. 
The proof of (\ref{eq:K135}) is as follows. 
In the integral in Eq.~(\ref{eq:K135}), we 
replace the two radial wavefunctions by their expressions (\ref{eq:K16}). 
Then, with the help of the formula \cite{PBM16} 
\begin{eqnarray}
\int_{0}^{\infty}{\rm e}^{-cx}x^{\alpha-1}
L_m^{\gamma}(cx)L_n^{\lambda}(cx)dx =
\nonumber
\\
\times \frac{(\gamma+1)_m (\lambda-\alpha+1)_n \Gamma(\alpha)}
{m!n!c^{\alpha}}
\nonumber
\\
\times {_3F_2}(-m,\alpha,\alpha-\lambda;\gamma+1,\alpha-\lambda-n;1),
\nonumber 
\end{eqnarray}
we arrive at
\begin{eqnarray}
J_{qq'}^{(\pm)} = \Omega \frac{\Gamma(q'+q+ c \pm b+1)}
{\Gamma(2q'+c \pm b+2)\Gamma(q-q'+1)}
\nonumber
\\
\times \sqrt{\frac{(n - q')!\Gamma(n + q'+ c \pm b+2)}
                  {(n - q )!\Gamma(n + q + c \pm b+2)}} 
\nonumber
\\
\times {_2F_1}(-q+q',q+q'+c \pm b+1;2q'+c \pm b+2;1).
\label{eq:K137}
\end{eqnarray}
By using the Gauss summation formula \cite{BE13}
\begin{eqnarray}
{_2F_1}(a,b;c;1)=\frac{\Gamma(c)\Gamma(c-a-b)}
{\Gamma(c-a)\Gamma(c-b)}
\nonumber
\end{eqnarray}
we can rewrite (\ref{eq:K137}) as
\begin{eqnarray}
J_{qq'}^{(\pm)} = \frac{\Omega}{q + q' + c \pm b + 1}
\nonumber
\\
\times \sqrt{\frac{(n - q')!\Gamma(n + q' + c \pm b + 2)}
                  {(n - q )!\Gamma(n + q  + c \pm b + 2)}}
\nonumber
\\
\times [\Gamma(q-q'+1)\Gamma(q'-q+1)]^{-1}.
\nonumber
\end{eqnarray}
This completes the proof of Eq.~(\ref{eq:K135}) since
$[\Gamma(q-q'+1)\Gamma(q'-q+1)]^{-1}=\delta_{qq'}$.

\vspace{0.3cm}
{\bf Appendix: The Smorodinsky-Winternitz and Morse Systems}
\vspace{0.1cm}

The Morse system with the potential
\begin{eqnarray}
V_{\rm M} = V_{0} ({\rm e}^{-2ax} -  2 {\rm e}^{-ax})
\nonumber
\end{eqnarray}
can be connected to the dynamical system with the potential
\begin{eqnarray}
V_{\rm SW} = \frac{\Omega^2}{2} z^2 + \frac{P}{2} \frac{1}{z^2}.
\nonumber 
\end{eqnarray}
(The latter potential may be considered 
as a one-dimensional component of the so-called 
Smorodinsky-Winternitz \cite{SW129,SW230,SW331} 
potential. The potential $V_{\rm SW}$ was investigated 
by Calogero \cite{CA32}.)

The Schr\"odinger equation for the Morse potential $V_{\rm M}$, i.e.,
\begin{eqnarray}
\left[ d_{xx} + 
2 E - 2 V_{0}({\rm e}^{-2ax} - 2 {\rm e}^{-ax})
\right] \psi = 0
\label{eq:K142}
\end{eqnarray}
admits a discrete spectrum (with $E < 0$) and a continuous spectrum.
For the discrete spectrum, by making the change of variable 
\begin{eqnarray}
y = a x,                         \,\,\,\, 
y \in {\bf R},                   \,\,\,\,
z = {\rm e}^{-s y},              \,\,\,\, 
z \in {\bf R}^{+}                
\nonumber
\end{eqnarray}
and the change of function
\begin{eqnarray}
\psi(x) = \frac{1}{\sqrt z} f(z) 
\nonumber
\end{eqnarray}
in Eq.~(\ref{eq:K142}), we get 
\begin{eqnarray}
\left[ d_{zz} + 4 \lambda^2 (2 - z^2) 
              + \left( \frac{8E}{a^2} + \frac{1}{4} \right) \frac{1}{z^2}
\right] f = 0, 
\label{eq:K145}
\end{eqnarray}
where 
\begin{eqnarray}
\lambda = \frac{\sqrt{2V_0}}{a}.
\nonumber
\end{eqnarray}
Equation (\ref{eq:K145}) has the same form as Eq.~(\ref{eq:K38bis}) 
for $z > 0$ with 
\begin{eqnarray}
E_{z} = 4 \lambda^2, \,\,\,\,\,\,
 \Omega = 2 \lambda, \,\,\,\,\,\,
P = - \frac{8 E}{a^2} - \frac{1}{4}.
\nonumber
\end{eqnarray}
Therefore, we must consider two admissible regions for the 
energy $E$: (i)     $    - 32 E > a^2$ and 
            (ii)    $0 < - 32 E < a^2 $. 

In the case (i), by employing the energy formula (\ref{eq:K45}) 
for $E_{z}$, we obtain that $E$ is determined by the relation
\begin{eqnarray}
\frac{\sqrt{- 2 E  }}{a} = \lambda - \left( p + s \right), 
\quad p = 0, 1, \cdots, \left[ \lambda - s \right]. 
\label{eq:K147} 
\end{eqnarray}
In Eq.~(\ref{eq:K147}), 
$[x]$ stands for the integral value of $x$. As a result, we have
\begin{eqnarray}
E = - V_{0} \left[ 1 - \frac{1}{\lambda} \left(p + s\right) \right]^2, 
\quad p = 0, 1, \cdots, \left[ \lambda - s \right]. 
\label{eq:K148} 
\end{eqnarray}
Equation (\ref{eq:K148}) 
is in agreement with the well-known result according to which
the discrete spectrum of the Morse system has a finite number
(here $[ \lambda - s ] + 1$) 
of energy levels with the condition $\lambda > s$. 

In the case (ii), we have 
\begin{eqnarray}
\frac{\sqrt{- 2    E}}{a} = \pm [ \lambda - \left( p + s \right) ],
\nonumber
\end{eqnarray}
which has no solution for $p \in {\bf N}$.
  
The connection just described between the Morse and 
Smorodinsky-Winternitz systems can be used also to 
deduce the wavefunctions of one system from the 
wavefunctions of the other. For instance, from 
Eq.~(\ref{eq:K42}), we immediately get the normalized 
solution $\psi (x) \equiv {\psi}_p (z; \lambda)$
of (\ref{eq:K142}):
\begin{eqnarray}
{\psi}_p (z; \lambda) 
= (-1)^{p} {(2\lambda)}^{\lambda - p}
\sqrt{ \frac{a p!}{\Gamma(2\lambda -p)} }
\nonumber
\\
\times 
{\rm e}^{- \lambda z^2} z^{2\lambda - 2p - 1}
                      L_p^{2\lambda - 2p - 1}(2\lambda z^2),
\label{eq:K150}
\end{eqnarray}
with
\begin{eqnarray}
z = {\rm e}^{-s a x}, \,\,\,\,\,\, p = 0, 1, \cdots, [ \lambda - s ].
\nonumber 
\end{eqnarray}
Our result (\ref{eq:K150}) differs from the one of 
Nieto and Simmons \cite{NS33} (by the fact that the 
factor $p!$ in (\ref{eq:K150}) is $2p-\lambda$ in 
Ref.~\cite{NS33}). 


\end{document}